\documentclass[twocolumn]{aastex63}
\usepackage{times}
\usepackage{amsmath}
\usepackage{textcomp}
\usepackage{amssymb}
\usepackage{natbib}
\usepackage{float}
\usepackage{color}
\usepackage{graphicx}

\newcommand{\lum}{erg\,s$^{-1}$}
\newcommand{\fermi}{{\it Fermi}}

\newcommand{\phflux}{\mbox{${\rm \, ph \,\, cm^{-2} \, s^{-1}}$}}

\newcommand{\gm}{$\gamma$}
\newcommand{\hbeta}{H{$\beta$}}
\newcommand{\halpha}{H{$\alpha$}}

\submitjournal{ApJ}

\shorttitle{S5 1027+74}
\shortauthors{Paliya et al.}

\begin{document}
\title{The Detection of Teraelectronvolt Radiation from a Flat Spectrum Radio Quasar}

\correspondingauthor{Vaidehi S. Paliya}
\email{vaidehi.s.paliya@gmail.com}

\author[0000-0001-7774-5308]{Vaidehi S. Paliya}
\affiliation{Inter-University Centre for Astronomy and Astrophysics (IUCAA), SPPU Campus, 411007, Pune, India}

\author[0000-0002-8434-5692]{M. B\"ottcher}
\affiliation{Centre for Space Research, North-West University, Potchefstroom, 2531, South Africa} 

\author[0000-0003-0841-7823]{Kiran Wani}
\affiliation{Indian Institute of Astrophysics, Block II, Koramangala,Bengaluru, Karnataka 560034, India}

\author[0000-0003-0545-5998]{P. N. Naseef Mohammed}
\affiliation{Farook College, Calicut, Kerala, 673632, India}

\author[0000-0002-4998-1861]{C. S. Stalin}
\affiliation{Indian Institute of Astrophysics, Block II, Koramangala,Bengaluru, Karnataka 560034, India}

\author[0009-0009-8004-8314]{S. Sahayanathan}
\affiliation{Astrophysical Sciences Division, Bhabha Atomic Research Centre, Mumbai-400085, India}
\altaffiliation{Homi Bhabha National Institute, Mumbai-400094, India}

\author[0000-0002-4464-8023]{D. J. Saikia}
\affiliation{Inter-University Centre for Astronomy and Astrophysics (IUCAA), SPPU Campus, 411007, Pune, India}

\author[0000-0002-4024-956X]{S. Muneer}
\affiliation{Indian Institute of Astrophysics, Block II, Koramangala,Bengaluru, Karnataka 560034, India}

\begin{abstract}
The very high-energy (VHE; $>$100 GeV) radiation carries the signatures of the matter-energy interaction in some of the most extreme astrophysical environments. Considering broad emission line blazars, i.e., flat spectrum radio quasars (FSRQs), the dense photon fields surrounding the relativistic jet can prohibit the particle population from accelerating to very high energies and producing VHE radiation. They can also possibly make the environment opaque for the VHE \gm~rays due to \gm\gm~pair production, thus explaining the paucity of VHE-detected FSRQs and non-detection of TeV radiation ($>$1 TeV) from them. Here we report, for the first time, a $>$7$\sigma$ detection of an FSRQ, S5 1027+74 ($z=0.123$), in the VHE band, including the first ever detection of TeV emission from an object of this class, using the Fermi Large Area Telescope observations. Its \gm-ray spectrum covering the 100 MeV to 2 TeV band revealed a prominent spectral break with a flat, rising shape above $\sim$10 GeV, a feature never detected from other VHE-detected FSRQs. The radio-to-\gm-ray spectral energy distribution of S5 1027+74 provides strong evidence of a third bump peaking at multi-TeV energies. These enigmatic findings imply that FSRQ jets can accelerate particles to extremely high energies and provide tantalizing clues about the complex radiative environment of relativistic jets.

\end{abstract}


\section{Introduction}{\label{sec:Intro}}
Relativistic jets are the cosmic laboratories to explore particle acceleration in extreme environments. Active Galactic Nuclei (AGN) hosting closely aligned relativistic jets are called blazars. They are classified as flat-spectrum radio quasars (FSRQs) and BL Lac objects with the former exhibiting broad emission lines in their optical spectra \citep[rest-frame equivalent width $\gtrsim5$\AA; cf.][]{2012ApJ...748...49S}. A physically motivated blazar classification has also been proposed according to which FSRQs are powered by radiatively efficient accretion \citep[$L_{\rm acc}/L_{\rm Edd}>1\%$;][]{2011MNRAS.414.2674G}. Due to the presence of an intense accretion disk, broad line region (BLR), and dusty torus photons surrounding the jet in FSRQs, the relativistic electrons rapidly cool down via inverse Compton scattering before attaining very high energies \citep[e.g.,][]{2009ApJ...692...32D}. This hypothesis explains the low-frequency peaked and Compton-dominated nature of the spectral energy distribution (SED) of FSRQs \citep[e.g.,][]{1998MNRAS.301..451G}. These objects usually have the high-energy hump peaking at $\sim$MeV energies and exhibit a soft \gm-ray spectrum above 100 MeV as observed by the Fermi Large Area Telescope \citep[LAT;][]{2009ApJ...697.1071A}. In the case of BL Lac objects, on the other hand, the photon fields external to the jet are weak or absent due to a radiatively inefficient accretion flow. Therefore, the radio-to-\gm-ray SEDs of these sources exhibit relatively high-frequency peaked humps since the jet electrons can accelerate to very high energies before cooling via radiative losses. The observation of a hard \gm-ray spectrum from BL Lac sources in the Fermi-LAT energy band supports this hypothesis \citep[cf.][]{2022ApJS..263...24A}.

At very high energies (VHE, $>$100 GeV), the known source population is dominated by BL Lac objects\footnote{\url{https://tevcat.org}} \citep[][]{2008ICRC....3.1341W}. Several BL Lacs have also been detected at even higher energies, i.e., in the TeV band \citep[$>$1 TeV; e.g.,][]{2024ApJS..271...25C}. Contrastingly, only 10 FSRQs have been reported as VHE emitting sources, but none in the TeV band \citep[e.g.,][]{2011ApJ...730L...8A}. This is likely because these objects have a strong BLR photon field, which can severely absorb the VHE emission via pair production, so the emission region must be outside the BLR. More importantly, FSRQs usually have low-frequency peaked SEDs, implying the paucity of energetic particles needed to radiate VHE emission. Indeed, all known VHE-emitting FSRQs, except PKS 1510$-$089\footnote{PKS 1510$-$089 has also been detected in the VHE band during low \gm-ray flux activity periods \citep[][]{2018A&A...619A.159M}.}, have been detected in the VHE band during elevated \gm-ray activity episodes along with the shift of the SED peaks to high energies, suggesting the fresh injection of energetic particles to the emission region \citep[see, e.g.,][]{2021A&A...647A.163M}.

In this letter, we report, for the first time, the discovery of VHE and TeV \gm-ray emission from the blazar S5 1027+74 using Fermi-LAT observations. This object has been classified as a BL Lac in the Fermi-LAT \gm-ray source catalogs \citep[e.g.,][]{2022ApJS..260...53A}. Interestingly, its optical spectrum published by \citet[][]{1993A&AS..100..395S} revealed broad emission lines indicating it to be rather an FSRQ at redshift $z=0.123$. Therefore, this object not only belongs to the rare population of the VHE emitting FSRQs but also is the first source of this class from which TeV emission has been published.

\section{Fermi-LAT Data Reduction}\label{sec2}
We considered $\sim$16 years of the Fermi-LAT data covering the energy range of 100 GeV to 2 TeV and time interval MJD 54683-60587 (2008 August 4 to 2024 October 4). In a circular region of interest (ROI) of radius 5$^{\circ}$, only events belonging to class {\tt P8R3\_SOURCEVETO} (evclass=2048) were adopted. This event class has the same background rate as the {\tt SOURCE} class up to 10 GeV while, above 50 GeV, similar to the {\tt ULTRACLEANVETO} class but with 15\% more acceptance. A cut on the maximum zenith angle was set to 105$^{\circ}$, and we considered both {\tt FRONT} and {\tt BACK} converting events (evtype=3). To model the \gm-ray sky, we considered all sources lying within 7$^{\circ}$ from the target and included in the fourth data release of the fourth Fermi-LAT \gm-ray source catalog \citep[4FGL-DR4;][]{2022ApJS..260...53A,2023arXiv230712546B}. The model also included the recommended isotropic and Galactic diffuse emission templates \citep[][]{2016ApJS..223...26A}. We first optimized the ROI spectral parameters and estimated the detection significance in terms of the test statistic (TS). All sources with TS$<$1 were removed from the model. The final likelihood fitting was performed by varying the spectral parameters of all remaining sources. To derive the \gm-ray spectral parameters corrected for extragalactic background light (EBL) attenuation, we adopted the EBL model of \citet[][]{2011MNRAS.410.2556D}. Next, we employed the tool {\tt gtsrcprob} to estimate the probability of VHE events originating from the direction of S5 1027+74. 

We also analyzed the LAT data covering the 100 MeV to 2 TeV energy range to generate the mission-averaged, broadband \gm-ray spectrum and monthly binned light curve. We adopted {\tt SOURCE} class events and a circular ROI of 15$^{\circ}$ radius. All 4FGL-DR4 sources within 20$^{\circ}$ from S5 1027+74 were considered during the likelihood fitting. For the component-wise point spread function (PSF) analysis, we used the summed likelihood components adopted in the 4FGL catalog \citep[][]{2020ApJS..247...33A}. To determine the presence of a spectral break, we also fitted the source spectrum with a broken power law model and estimated the test statistic curvature \citep[TS$_{\rm curve;~BPL-PL}=2(\log\mathcal{L({\rm BPL})-\log L ({\rm PL})}$), where $\mathcal{L}$ is the likelihood function;][]{2012ApJS..199...31N}. Finally, light curves were generated by computing the TS of each source in each time bin and varying the spectral parameters of those having TS$>$25. In the case of non-convergence of the likelihood fitting, we also froze the spectral shapes and allowed only the normalization parameter to vary. The spectral parameters of the remaining sources were fixed to their mission-averaged value. The entire data analysis was performed using fermiPy \citep[][]{2017arXiv170709551W}.

\section{Results}\label{sec3}
\subsection{S5 1027+74: A TeV Blazar}
The \gm-ray data analysis covering more than 16 years of the Fermi-LAT operation led to a significant detection ($>7\sigma$) of S5 1027+74 above 100 GeV.  This is the first report of identifying a new VHE-emitting FSRQ. The derived spectral parameters are provided in Table~\ref{tab1}. We repeated the analysis considering {\tt SOURCE} class events and found similar results albeit with a slightly lower TS (Table~\ref{tab1}), which is possibly due to larger background rate of the {\tt SOURCE} class.

\begin{table*}
\caption{The results of the \gm-ray spectral analysis done in the energy range 0.1$-$2 TeV.}\label{tab1}
\begin{tabular}{lcccccc}
\hline
\hline
{\tt evclass} & $F_{\rm obs}$  & $\Gamma_{\rm obs}$ & TS$_{\rm obs}$ & $F_{\rm EBL}$ & $\Gamma_{\rm EBL}$ & TS$_{\rm EBL}$ \\
\hline
{\tt SOURCEVETO} & $7.1\pm3.0$ & $1.6\pm0.5$ & 51 & $7.1\pm3.0$ & $1.0\pm0.6$ & 51\\
{\tt SOURCE} & $5.8\pm2.6$ & $1.5\pm0.6$ & 38 & $5.8\pm2.6$ & $0.9\pm0.6$ & 38\\
\hline
 & & VHE photons & & & & \\
Energy & MET  & R.A. (J2000) & Decl. (J2000) & Ang. sep. & Asso. Prob. & Asso. Prob.\\
(GeV)  & (UT) & (degrees)    & (degrees)     & (degrees) & ({\tt SOURCEVETO}) & ({\tt SOURCE}) \\
\hline
1304.2 & 638615791.48 & 157.569 & 74.663     &     0.081               & 0.996 & 0.968\\
       & (2021 Mar 28 09:16:26.48) &  & & & & \\
341.8  & 456476800.99 & 157.822 & 74.725    &      0.026              & 0.999 & 0.996\\
       & (2015 Jun 20 07:06:37.99) &  & & & & \\
259.1  & 553970897.08 & 157.983 & 74.670    &      0.048              & 0.994 & 0.987\\
       & (2018 Jul 22 16:48:12.08) &  & & & & \\
128.5  & 634266998.20 & 157.840 & 74.784    &      0.084              & 0.970 & 0.956\\
       & (2021 Feb 06 01:16:33.20) &  & & & & \\
\hline
\end{tabular}
\tablecomments{The quoted flux values (columns 2 and 5) are in units of 10$^{-12}$ \phflux. The parameters with subscript `obs' and `EBL' refer to observed and EBL-attenuation corrected quantities, respectively. The angular separation was estimated using the radio coordinates of S5 1027+74 \citep[R.A. = 157.841767, Decl.=74.699540, J2000;][]{1991ApJS...75....1B}. The association probability that the event belongs to S5 1027+74 was computed using the tool {\tt gtsrcprob}.
}
\end{table*}

\begin{figure}
\includegraphics[width=\linewidth]{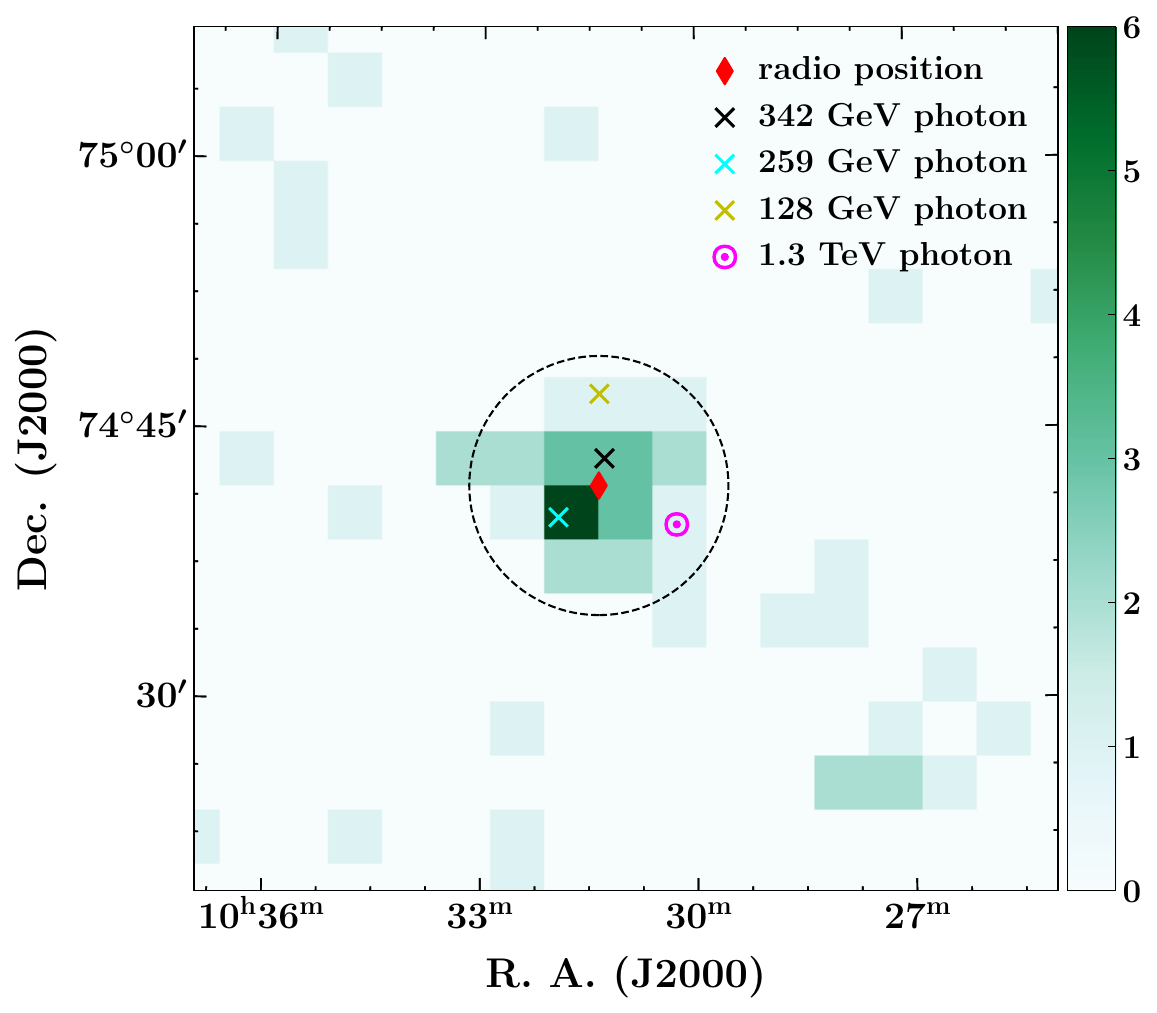}
\caption{The 10-2000 GeV Fermi-LAT counts map (scale = 0$^{\circ}$.05 pixel$^{-1}$) centered at the radio position of S5 1027+74. The dotted circle refers to 68\% Fermi-LAT PSF containment radius above 100 GeV for front-converting events ($\sim$0$^{\circ}$.1). The positions of $>$100 GeV events estimated by {\tt gtsrcprob} tool are highlighted with different markers as labeled.
} \label{fig:1}
\end{figure}

The {\tt gtsrcprob} tool revealed the detection of four VHE photons with the association probability $>$95\% (Table~\ref{tab1}). We show the count map of {\tt P8R3\_SOURCEVETO} events with 10$-$2000 GeV energies around S5 1027+74 in Figure~\ref{fig:1} and highlighted the positions of the VHE photons lying within the 68\% containment radius of the Fermi-LAT PSF above 100 GeV \citep[$\sim0^{\circ}.1$;][]{2021ApJS..256...12A}. Using the association probabilities estimated by {\tt gtsrcprob} and applying Fisher's method, we derived the null hypothesis probability of $1.3\times10^{-6}$ that the detected VHE events belong to the diffuse foreground/background or other 4FGL sources. This allowed us to reject the null hypothesis probability at 4.7$\sigma$ confidence level, ensuring a reliable association with S5 1027+74. 

Interestingly, one of the VHE photons had energy $\sim$1.3 TeV, suggesting the source to be a TeV emitter. We further ascertained the robustness of the TeV detection by repeating the Fermi-LAT data analysis with {\tt SOURCE} class events and found similar results though with marginally reduced association probability, likely due to increased background rate of this event class. This TeV photon-pair conversion took place in the front conversion layer of the LAT tracker (convtype=0) and has intrinsically better angular resolution than those that convert in the back section, thus robustly confirming its association. This finding makes S5 1027+74 the only FSRQ from which TeV emission has been published. This interesting result is further supported by the observed spectral parameters described below.

\begin{figure*}
\hbox{
\includegraphics[scale=0.38]{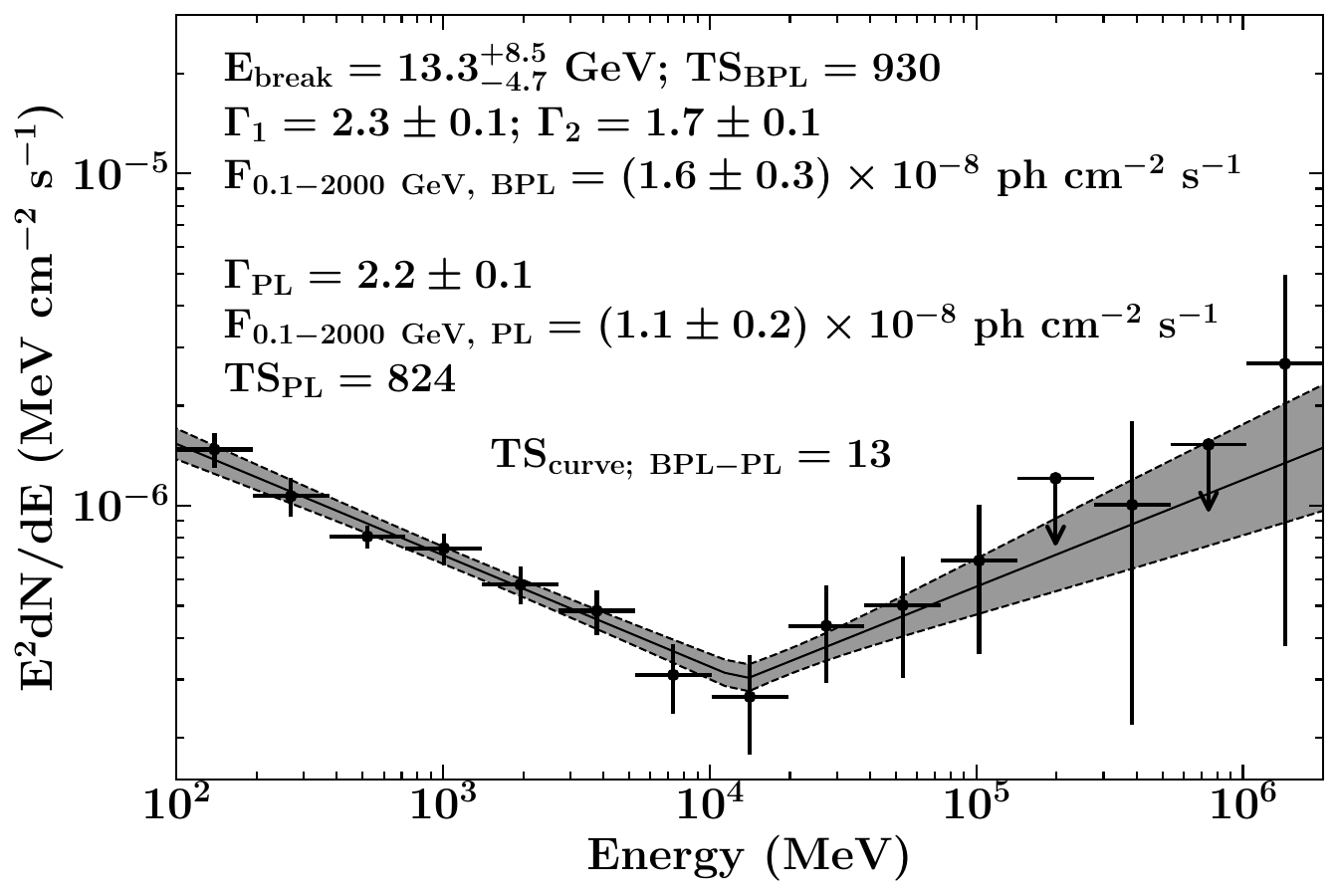}
\includegraphics[scale=0.42]{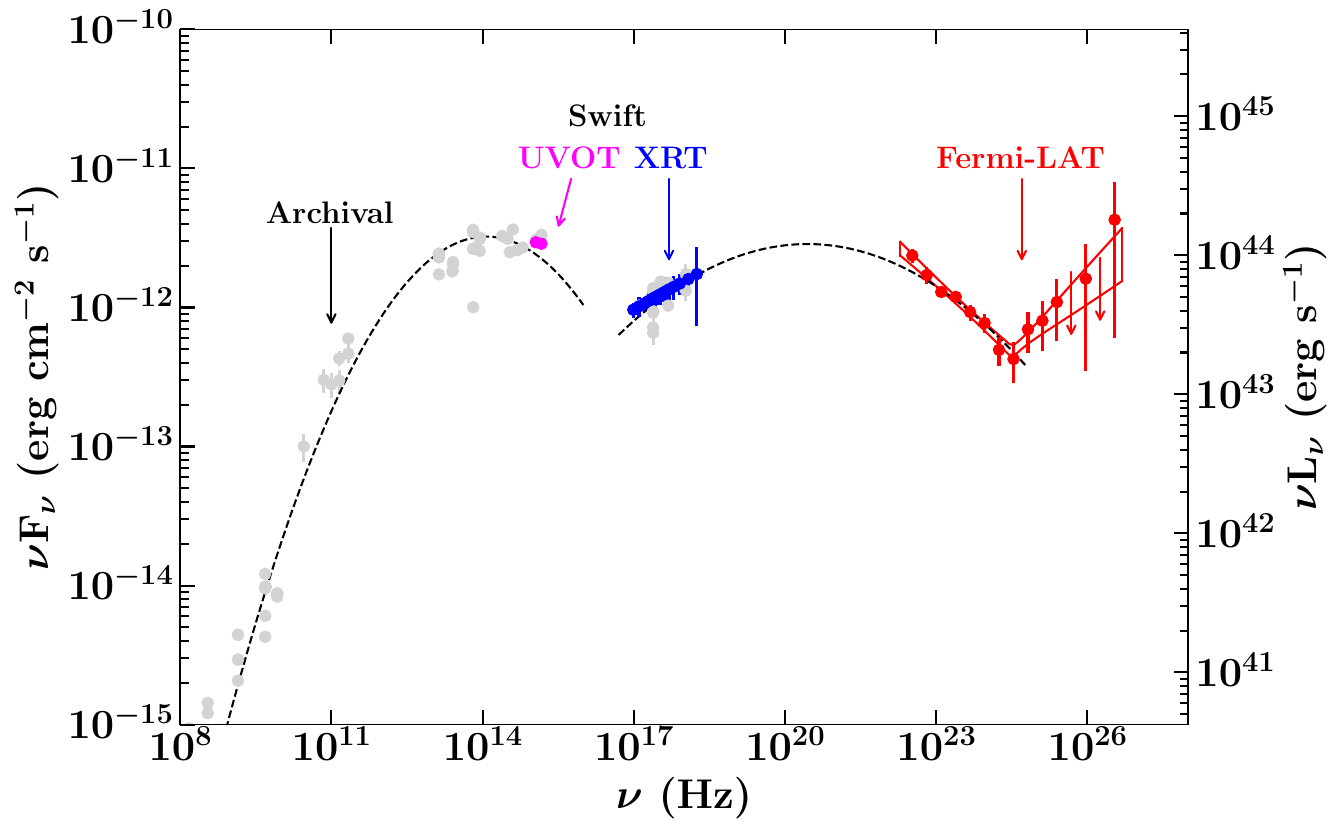}
}
\hspace{2.0cm}
\vbox{
\includegraphics[scale=0.37]{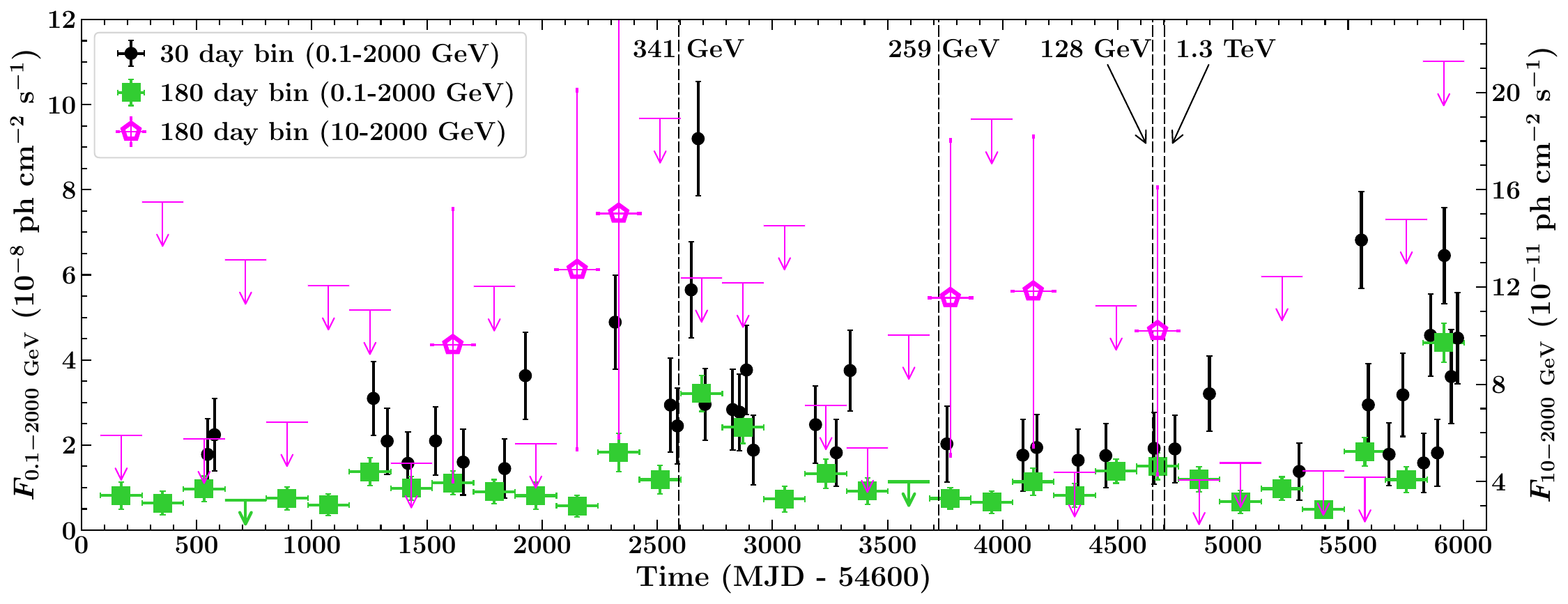}
\includegraphics[scale=0.37]{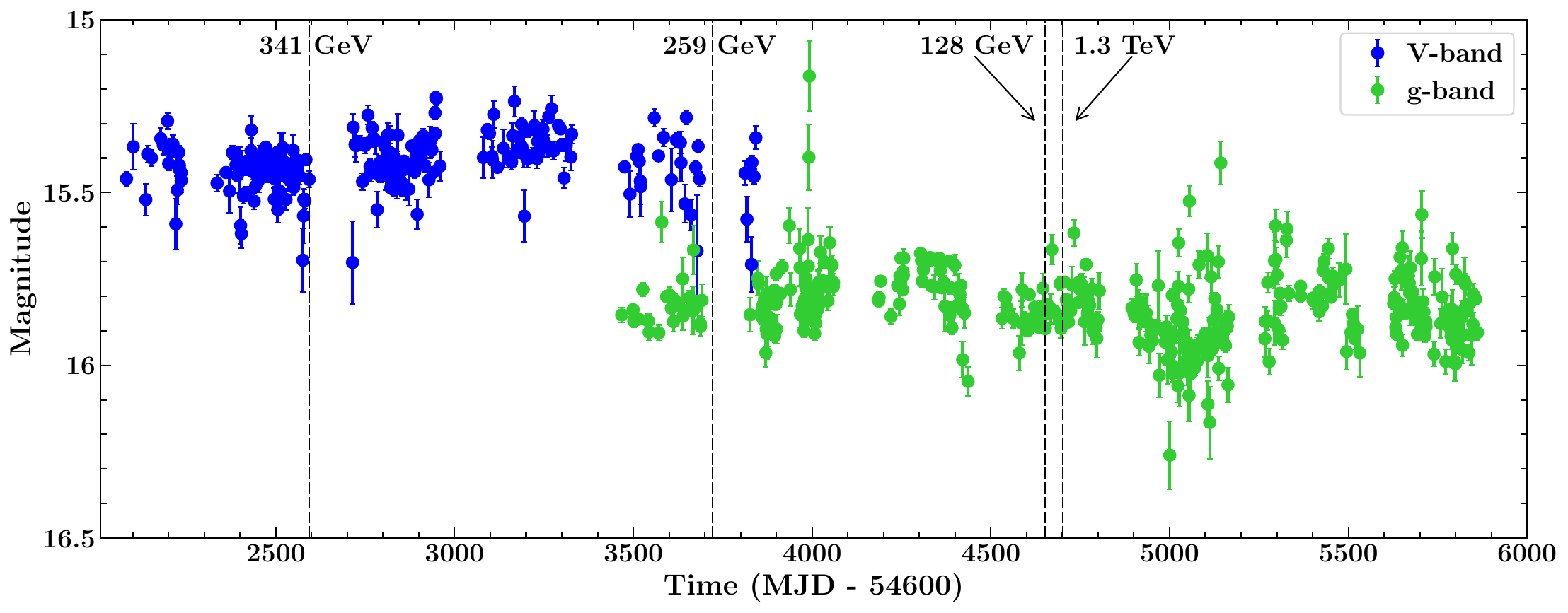}
}
\caption{Top left: The \gm-ray spectrum of S5 1027+74. The power law and broken power law model parameters are quoted. Top right: The broadband SED along with the fitted polynomial function (dashed line). Middle: The \gm-ray light curves generated using different binning are shown in this panel. The arrival times of VHE photons are highlighted by vertical dashed lines. Flux upper limits are at estimated 95\% confidence level. Bottom: The $V$- and $g$-bands light curve using ASAS-SN observations.} \label{fig:2}
\end{figure*}

A noteworthy finding of the Fermi-LAT observations is the 0.1$-$2 TeV \gm-ray spectral shape, which was extremely flat with a derived photon index of $\Gamma_{\rm 0.1-2~TeV}=1.6\pm0.5$. In contrast, the VHE spectra of all other FSRQs detected above 100 GeV were found to be soft \citep[$\Gamma_{\rm >0.1~TeV}>3$; see, e.g.,][]{2008Sci...320.1752M,2015ApJ...815L..22A,2016A&A...595A..98A,2020A&A...633A.162H}. The EBL-attenuation corrected \gm-ray spectrum appears even harder. A hard \gm-ray spectrum of S5 1027+74 above 10 GeV was also reported in the third catalog of hard Fermi-LAT sources \citep[3FHL, $\Gamma_{\rm 0.01-2~TeV}=1.70\pm0.22$;][]{2017ApJS..232...18A}. To explore further, we generated the full 0.1$-$2000 GeV \gm-ray spectrum using $\sim$16 years of Fermi-LAT data and show it in the top left panel of Figure~\ref{fig:2}. A broken power law model was preferred over a power law model at $\sim$3$\sigma$ confidence level (TS$_{\rm curve}=13$)\footnote{The TS$_{\rm curve}$ was estimated from the derived log-likelihood values for the broken power law ($\log\mathcal{L}=549292.6$) and power law ($\log\mathcal{L}=549286.1$) model fitting.}. The break energy was estimated to be $E_{\rm break}=13.3^{+8.5}_{-4.7}$ GeV. The \gm-ray spectrum above $E_{\rm break}$ has a rising shape in the $\nu F_{\nu}$ versus $\nu$ plane and is consistent with the observed VHE spectral shape. Such inverted spectral features have also been systematically searched for in high-synchrotron peak BL Lac objects \citep[][]{2025arXiv250702718D}.

The arrival times of the VHE photons are highlighted in the \gm-ray light curve of S5 1027+74 shown in the middle panel of Figure~\ref{fig:2}. We also show the optical flux variations using $V$- and $g$-band observations taken with the All-Sky Automated Survey for SuperNovae \citep[ASAS-SN;][]{2014ApJ...788...48S,2023arXiv230403791H}. None of the VHE detection epochs coincided with any elevated \gm-ray or optical flux activity periods, implying that the VHE emission originated during quiescent states. This finding makes S5 1027+74 only the second FSRQ detected in the VHE band during the low activity state.

The Neil Gehrels Swift satellite observed S5 1027+74 thrice during 2010-2011 when the blazar was in quiescence. Using these observations and the mission-averaged Fermi-LAT spectrum, we generated the radio-to-\gm-ray SED (Figure~\ref{fig:2}, top right panel). By fitting a second-order polynomial function, we estimated the synchrotron and inverse Compton peak frequencies to be $1\times10^{14}$ Hz and $3\times10^{20}$ Hz, respectively. Interestingly, the rising \gm-ray spectrum above $E_{\rm break}$ hints at a third SED peak beyond 2 TeV. To our knowledge, such an intriguing feature has never been published for any FSRQ.

\subsection{S5 1027+74: An FSRQ}
\begin{figure}
\vbox{
\includegraphics[width=\linewidth]{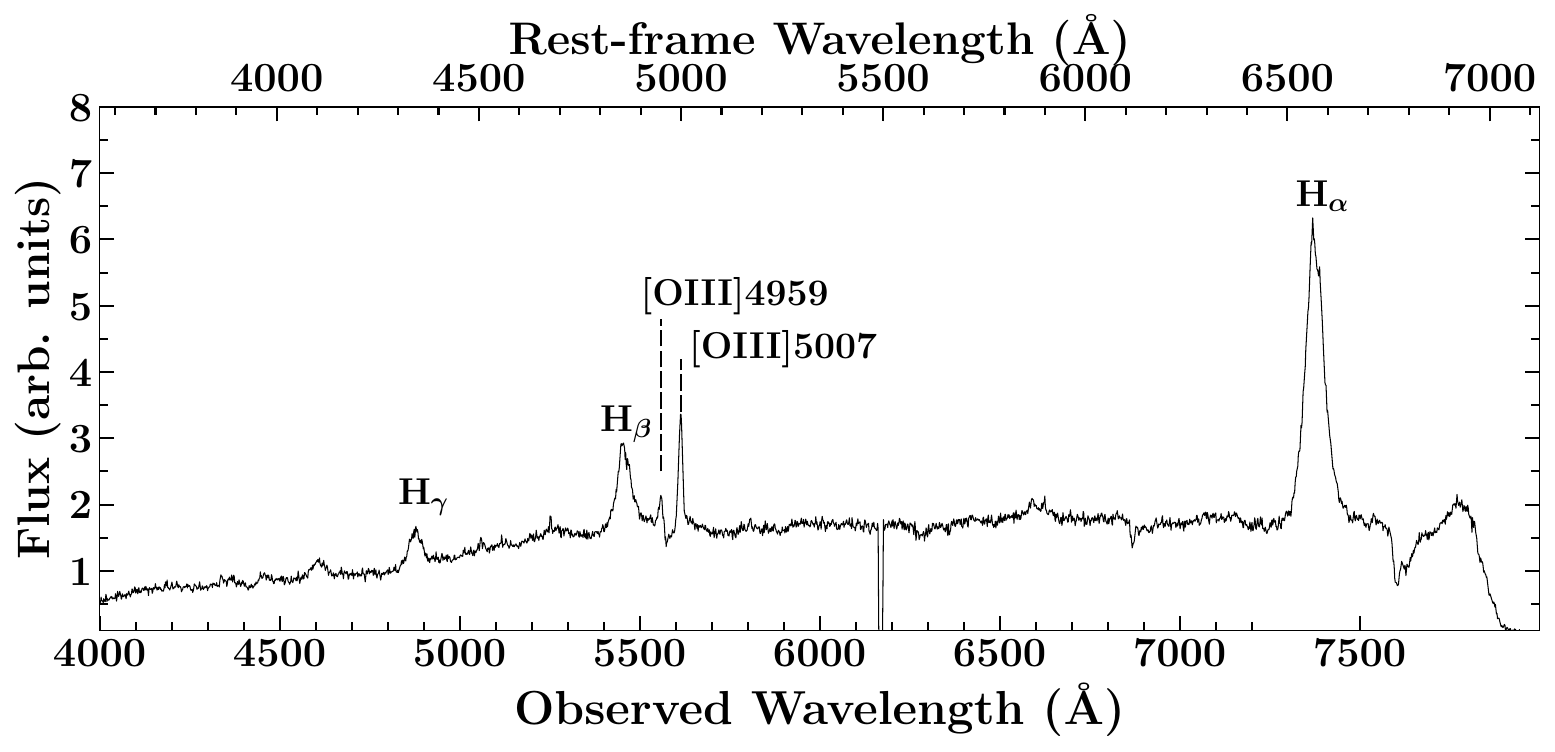}
}
\caption{Optical spectrum of S5 1027+74 taken on 2025 April 17 with Himalayan Chandra Telescope.} \label{fig:3}
\end{figure}

Given the detection of TeV emission and a possible FSRQ nature, we procured a new optical spectrum of S5 1027+74 on 2025 April 17 with the Hanle Faint Object Spectrograph Camera mounted on the 2-m Himalayan Chandra Telescope with 3600 seconds of exposure. We adopted the grism-7, which provides a spectral coverage of 380$-$800 nm. The data reduction was performed following the standard procedures \citep[see, e.g.,][]{2022MNRAS.516.5702O}. In Figure~\ref{fig:3}, we show the wavelength-calibrated spectrum of the source. The broad emission lines \halpha~and \hbeta~were well detected (signal-to-noise ratio $>$10$\sigma$) which is similar to the earlier spectrum published by \citet[][]{1993A&AS..100..395S}. The rest-frame equivalent widths of the broad \halpha~and \hbeta~emission lines were reported to be 124 \AA~and 85 \AA, respectively \citep[][]{2021ApJS..253...46P}. These results unambiguously confirm the FSRQ nature of S5 1027+74.

\section{Discussion}\label{sec4}
\subsection{VHE and TeV Emission -- Constraints on the Emission Region Location}
In FSRQs, intense BLR radiation that can provide seed photons for inverse Compton scattering can also absorb \gm-ray photons via pair production \citep[e.g.,][]{2006ApJ...653.1089L,2016ApJ...821..102B}. Moreover, the inverse Compton scattering cross-section for BLR (dominated by hydrogen Lyman $\alpha$ photons) and dusty torus or DT (characteristic frequency of $\sim 3 \times 10^{13}$~Hz) photon fields also decreases beyond characteristic energies of
\begin{eqnarray}
    E_{\rm KN,~BLR} &= 15\frac{\delta}{\Gamma (1+z)}~{\rm GeV}, \cr 
    E_{\rm KN,~DT} &= 1.2\frac{\delta}{\Gamma (1+z)}~{\rm TeV}, \cr
\label{KNconstraints}
\end{eqnarray}
respectively, due to Klein-Nishina effects \citep[cf.][]{2009MNRAS.397..985G}.
For S5 1027+74, these energies are $E_{\rm KN,~BLR} \sim$13~GeV and $E_{\rm KN,~DT} \sim$1~TeV. This suggests that the VHE emitting region may be located outside the BLR, where the primary seed photons for the inverse Compton scattering are provided by the dusty torus. 

\begin{figure}
\hbox{\hspace{-0.4cm}
\includegraphics[scale=0.35]{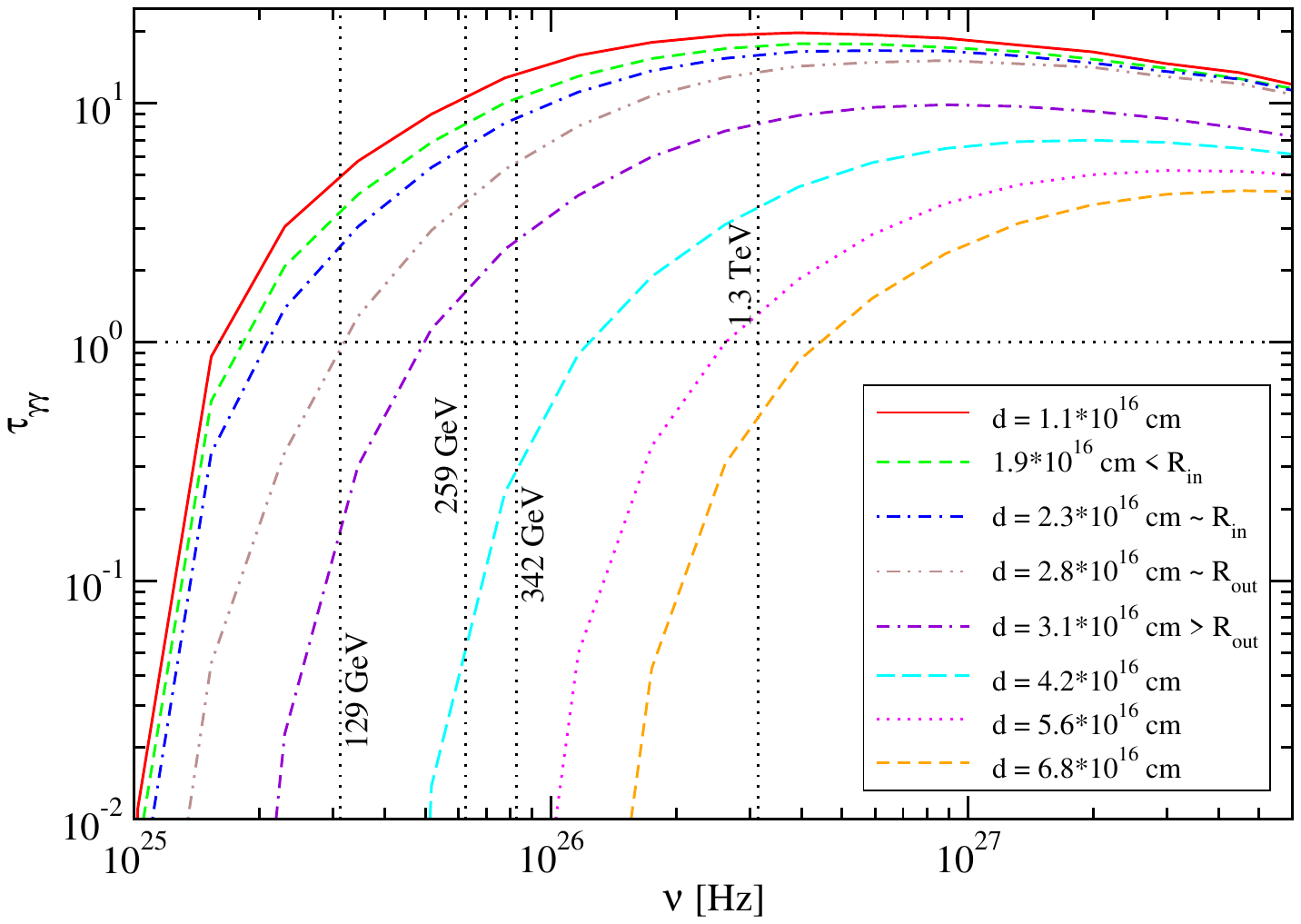}
}
\caption{Optical depth due to \gm\gm~pair production on BLR photons as a function of the \gm-ray photon frequency for different locations of the $\gamma$-ray emission region along the jet axis.} \label{fig:4}
\end{figure}

In order to quantify constraints due to BLR $\gamma\gamma$ pair production opacity we used the prescription of \citet[][]{2016ApJ...821..102B,2018ApJ...869...87B}. Specifically, the BLR optical depth was estimated as a function of distance ($d$) of the emission region from the central black hole and photon energy. 
We adopted an accretion disk luminosity of $L_{\rm disk} = 7.6 \times 10^{43}$ \lum, and assumed the BLR luminosity to be 10\% of $L_{\rm disk}$ \citep[][]{2021ApJS..253...46P}. The BLR was modeled as a homogeneous shell with inner and outer radii of $R_{\rm in} = 0.9 R_{\rm BLR}$ to $R_{\rm out} = 1.1 R_{\rm BLR}$ where $R_{\rm BLR} = 2.6 \times 10^{16}$~cm \citep[][]{2016ApJ...821..102B}, and the BLR spectrum is modeled as the sum of the brightest quasar emission lines following the relative line luminosities of \cite{Francis91}. We show the derived optical depth as a function of photon energy for several representative distances $d$ in Figure~\ref{fig:4}. The energies of the detected VHE photons are highlighted by the vertical dotted lines. We conclude that, in order to avoid BLR absorption, the TeV photon has likely been emitted at $d \gtrsim 3 \times R_{\rm BLR}$.

\subsection{Third SED Bump}

The unusual, apparently two-component shape of the $\gamma$-ray SED of S5~1027+74 deserves special consideration. While in the framework of leptonic blazar emission models, the high-energy SEDs of FSRQs are usually well modeled with external-Compton (EC) scattering of BLR and DT photons --- EC(BLR) and EC(DT), respectively ---, the decline of the Klein-Nishina cross-section at energies above $\sim$1 TeV (see Equation~\ref{KNconstraints}), where the cross-section is already reduced to 1/3 of its value in the Thomson regime for EC(DT) make such an interpretation for the hard VHE $\gamma$-ray spectrum very unlikely. Reproducing the VHE spectrum as an additional EC component would require a combination of a very low-energy target photon field and extremely high-energy electrons. The Cosmic Microwave Background (CMB) could provide such a photon field, upscattered by multi-TeV electrons \citep{Boettcher08}. Given the similarity of the spectral indices of the X-ray spectrum (photon index $1.79 \pm 0.11$) and the VHE spectrum (photon index $1.68 \pm 0.12$), the same electron distribution could potentially be responsible for both emissions via synchrotron self Compton and EC(CMB) processes, respectively, while the GeV $\gamma$-ray emission could be produced by the more standard EC(BLR) or EC(DT) mechanisms. We note that the non-simultaneous nature of the Fermi-LAT and Swift-XRT datasets means a similarity of the X-ray and VHE photon indices should be considered with caution.

An alternative interpretation could be a lepto-hadronic model in which the GeV $\gamma$-ray emission is produced by leptonic EC processes, while the VHE $\gamma$-rays arise from synchrotron emission by ultra-relativistic protons. Such an interpretation of a VHE spectrum peaking at multi-TeV energies would require the acceleration of protons to extreme energies of $E_{\rm p, max} \sim 2 \times 10^{20} \, (\delta/10)^{-1/2} \, (B / {\rm G})^{-1/2}$~eV, i.e., in the realm of ultra-high-energy cosmic rays for moderate values of the Doppler factor $\delta$ and magnetic field $B$. 

Yet another possibility is the interpretation of the apparent two-component shape of the $\gamma$-ray spectrum as an absorption feature around the break energy of $13^{+8.5}_{-4.7}$~GeV. A $\gamma\gamma$ absorption feature due to BLR radiation dominated by Ly$\alpha$ photons for a source at redshift $z = 0.123$ would have its maximum depth at $\sim 22$~GeV, marginally consistent with the observed break energy. The underlying GeV -- TeV $\gamma$-ray continuum could then be produced either by leptonic SSC or EC(CMB) emission or by hadronic processes.

Though a detailed exploration of the scenarios discussed above will be presented in a forthcoming, dedicated publication, they imply the existence of complex radiative processes in relativistic jets and their possible interplay with the ambient environment. With the advent of Cherenkov Telescope Array, it will be possible to characterize the VHE radiation of a larger FSRQ population and delve deeper into the physics of relativistic jets.

During the review of the manuscript, we learned about the report of TeV emission from another FSRQ, 3C~273, presented in International Cosmic-Ray Conference 2025\footnote{\url{https://indico.cern.ch/event/1258933/contributions/6491204/attachments/3104111/5500883/Benbow\_ICRC2025.pdf}}. The detections of TeV emission from these two distinct FSRQs can be interpreted within the same scenarios.

\acknowledgements
We thank the journal referee for constructive criticism.
This research has made use of NASA’s Astrophysics Data System Bibliographic Services. The use of the \fermi-LAT data provided by the Fermi Science Support Center is gratefully acknowledged.

\software{fermiPy \citep{2017arXiv170709551W}}

\bibliography{Master}{}
\bibliographystyle{aasjournal}

\end{document}